\documentstyle[sprocl]{article}

\def\be{\begin{equation}}
\def\ee{\end{equation}}
\def\bea{\begin{eqnarray}}
\def\eea{\end{eqnarray}}

\begin{document}

\title{ Effective Hamiltonian, Mori Product and Quantum Dynamics}

\author{A.Cuccoli and V.Tognetti}

\address{Department of Physics, University and I.N.F.M. -  Florence, Italy}

\author{R.Giachetti}

\address{Department of Physics, University and I.N.F.N. -  Florence, Italy}

\author{R.Vaia}

\address{I.E.Q.-C.N.R. -  Florence, Italy}


\maketitle\abstracts{
An appropriate extension of the effective potential
  theory is presented that permits the approximate calculation of the dynamical
  correlation functions for quantum systems. These are obtained by
  evaluating the generating functionals of the Mori products of quantities related
  to the relaxation functions in the (PQSCHA) pure self consistent
  harmonic approximation.}

\section{Dynamic Correlations and Spectral Shapes}

The thermodynamics of quantum systems has been widely studied by
the effective potential theory \cite{GTall}: equilibrium properties
have been accurately determined \cite{CTVVmagall,CGTVV95}, but the
investigation of dynamical quantities is much more difficult . It
is our purpose to give some insights on how the dynamical problem
can be approached in terms of the Kubo relaxation functions, that
naturally appear in the framework of Mori theory \cite{Moriboth}.
These functions are obtained by suitably defined scalar products,
the Mori products
\begin{equation}
 R_{A,B}(t)=(\hat A|\hat B(t)) =
 \int_0^{\beta\hbar}\,du\,\langle\,\hat A(0)\,\hat B(t+iu)\,\rangle\,,\
 \label{1}
\end{equation}
where the observables $\hat{A}$ and $\hat{B}$ are taken such that
$\langle\hat{A}\rangle=\langle\hat{B}\rangle=0$, and braces denote
the thermodynamic average. The Laplace transform of Eq.(\ref{1})
and in particular, the self-relaxation function $\Xi_0(t)=(\hat
F_0|\hat F_0)^{-1}\,(\hat F_0(t)|\hat F_0)$ of an hermitian
operator $\hat F_0(t)$ can be Laplace transformed and expanded in a
continued fraction, namely
$\Xi_j(z)=(z+\delta_{j+1}\Xi_{j+1}(z))^{-1}$, where
$\delta_{j+1}=(\hat F_j|\hat F_j)^{-1}\,(\hat F_{j+1}|\hat
F_{j+1})$ and $\hat F_j$ denotes the so-called $j$-th fluctuating
force. The quantities $(\hat F_j|\hat F_j)$, can be related to a
combination of the first $2(j+1)$ moments of the time
series-expansion of $\Xi_0(t)$. While $(\hat F_j|\hat F_j)$ with
$j\not=0$ can be expressed in terms of static correlations of
time derivatives of $\hat F_0$, the
quantity $(\hat F_0|\hat F_0)$ requires the direct evaluation of
the Mori product.

Indeed, experiments measure the spectral shape, related to
$\Xi_0(z)$ by the ``detailed balance'' principle:
\begin{equation}
 {\cal S}(\omega)={(\hat F_0|\hat F_0)}\,\frac\omega{1-e^{-\beta
 \omega}}\,\frac 1\pi \,\Re\, (\Xi_0(z=i\omega)).
\end{equation}
Therefore, ${\cal S}(\omega)$ can be approached from the knowledge
of the static quantities $\delta_j$ up to a sufficiently large
number $j=J$ \cite{CMGTVall}, $\Xi_J(t)$
\cite{CGTVV95,LoveseyM72,CMGTVall}.

\section {Imaginary-time Ordered Products}

Here we shall provide a rigourous derivation of imaginary time
ordered products which allows us to control also the evaluation of
real time correlators.

We start from the generating functional in the hamiltonian
path-integral form:
\begin{eqnarray}
&Z&[L,J] = \oint{\cal D}[x(u)]\,\int{\cal D}[p(u)]\,\exp[-\frac
1\hbar\cal S]\cr
\cal S&=&\int_0^{\beta\hbar}du\Bigl
(-ip(u)\dot x(u)+{\cal H}\big(p(u),x(u)\big)-\hbar L(u)p(u) -
\hbar J(u)x(u)\,\Bigl)\,.
\end{eqnarray}

According to the effective potential method \cite{GTall,CGTVV95} we
use a quadratic trial action. The effective Hamiltonian is $\,
{\cal H}_{{}_{\rm
eff}}(\eta,\xi)=w(\eta,\xi)+\beta^{-1}\,\ln(f^{-1}\,\sinh f)~$,
where $f=\beta\hbar\omega/2$. By defining the two-component vectors
$\rho={}^t(\eta,\xi)$ and $K(u)={}^t(L(u),J(u))$, the approximated
generating functional can be written as:
\begin{eqnarray}
  Z_0[K]& =&\int
  \frac{d\eta\,d\xi}{2\pi\hbar}\,\exp{\displaystyle{-\beta{\cal
        H}_{{}_{\rm eff}}(\eta,\xi)}}\cr
        &\cdot&\exp\Bigl[\,
  \int_0^{\beta\hbar} du\,{}^t\!\rho \,K(u)+\frac {1}{2}
  \int_0^{\beta\hbar} du \int_0^{\beta\hbar} dv\,
  {}^t\!K(u)\,\Phi(u-v)\, K(v)\,\Bigr].
\label{zeta0}
\end{eqnarray}

In Eq.(\ref{zeta0}) we have introduced the $2\times 2$ matrix $\Phi_{k\ell}(u-v)$
with elements
$\Phi_{11}(u-v)=m^2\omega^2\,\Phi_{22}(u-v)=m^2\omega^2\Lambda_f(u-v)\,$
and $\Phi_{12}(u-v)=-\Phi_{21}(u-v)=\Gamma_f(u-v)\,,$ where
\begin{eqnarray}
  & &\Lambda_f(u-v)=\frac{\hbar}{2m\omega\sinh
    f}\Bigl[\,\cosh(\,|\omega(u-v)|-f\,)-\frac{\sinh f}f\,\Bigr] \cr &
  &\cr & &\Gamma_f(u-v)=-\frac{i \hbar}{2}\frac{\sinh(\,|\omega(u-v)|-f\,)}{\sinh
    f}\,\Bigl[\, \theta(u-v) -\,\theta(v-u)\,\Bigr]\,.
\end{eqnarray}
While $\Lambda_f$ is always well defined, the value of $\Gamma_f$
for $u=v$ is determined only when the limit $v-u\rightarrow 0\,^\pm$
is specified, reflecting the commutation relation of
$\hat x$ and $\hat p$ at the same time. Moreover we notice that,
$\Lambda_f(0)=\Lambda_f(\beta\hbar)=\alpha\,$ and that
$\Lambda_f(u-v)$ and $\Gamma_f(u-v)$ have a vanishing average in
$[0,\beta\hbar]$.

Defining the two-component vectors $\hat z={}^t( \hat p, \hat x)$
and $y=\big(y_1(u),y_2(u)\big)$, the following general formula can
be derived \cite{CGTV98} in the low-coupling approximation\nobreak
\cite{CGTVV95}.
\begin{equation}
 \Bigl\langle{{\cal T}_u}\Bigl[\prod_{\nu=1}^N \hat F_{\nu}( \hat
  z_{i_\nu}(u_\nu))\Bigr]\Bigr\rangle ={\cal N}\,\int
  \frac{d\eta\, d\xi}{2\pi\hbar}\, e^{\displaystyle{-\beta{\cal
        H}_{{}_{\rm eff}}(\eta,\xi)}} \Bigl\langle\Bigl\langle
  \prod_{\nu=1}^N
  F_{\nu}\big(\rho_{i_\nu}+y_{i_\nu}(u_\nu)\big)\Bigr\rangle\Bigr\rangle.
\label{genav}
\end{equation}
Where, ${\cal N}=Z^{-1}_0[0]$ is the normalizing factor. The
Gaussian average of the variables $y_{i_\nu}(u_\nu)$ is defined by
the second moments
\begin{equation}
\bigl\langle\bigl\langle y_i(v) y_j(u)
\bigr\rangle\bigr\rangle=\Phi_{ij}(u-v)\, .\label{gau}
\end{equation}
\noindent

\section {Discussion.}

The key result \cite{CGTV98} is represented by the expression
(\ref{genav}), with the definition (\ref{gau}). Indeed, complicated
static Mori products, {\it i.e.} moments of any order, can be
evaluated by this last equation. Static correlations can also be
obtained performing the appropriate limit $u-v\rightarrow 0$.

For instance, when $(u-v)\to 0^+$, we indeed recover our previous
result \cite{CTVVmagall,CGTVV95} for static averages:
\begin{equation}
  \langle \hat A\hat B\rangle= {\cal N}\,\int \frac{d\eta\,
    d\xi}{2\pi\hbar}\, \langle\langle
  AB\rangle\rangle\,e^{\displaystyle{-\beta {\cal H}_{{}_{\rm
          eff}}(\eta,\xi)}}\,,
\label{statAB}
\end{equation}

As far as the Mori product is concerned, the well known series
expansion $(2\pi\epsilon)^{-1/2}
\,\exp\{-x^2/(2\epsilon)\}=\sum_{n=0}^\infty\,(1/n!)\,
(\epsilon/2)^n\,\delta^{(2n)}(x)\,$ appears to be an efficient tool
to approximate the static Mori product of general operators, as
well as their dynamical correlations, when the scales of the
quantum fluctuations in the system, ruled by $\hbar$ and the
natural length scale $\alpha$, are small. It has to be noticed that
the averages of $\Lambda_f(u)$ and $\Gamma_f(u)$ in
$[0,\beta\hbar]$ are vanishing. An expansion in terms of these
quantities has been recently developed \cite{CGTV98}.

At the lowest order $(\hat A(p,x)|\hat B(p,x))$ reduces to the
``classical like'' average of the product of the Gaussian spreads
of the two operators taken at the same order:
\begin{equation}
  (\hat A(p,x)|\hat B(p,x)) ={\cal N}\,\beta\hbar\, \int \frac{d\eta\,
    d\xi}{2\pi\hbar}\, \langle\langle
  A(\eta,\xi)\rangle\rangle\,\langle\langle
  B(\eta,\xi)\rangle\rangle\, e^{\displaystyle{-\beta{\cal H}_{{}_{\rm
          eff}}(\eta,\xi)}}\,+o(\alpha,\hbar)\, .
\label{morexpsimpl}
\end{equation}
This zeroth order coincides with the assumption proposed in an
attempt \cite{CaoVall} to approach also the quantum correlators.

In order to do this, let us notice that harmonic oscillators evolve
by the same law both in classical and in quantum dynamics: the
differences between quantum and classical statistical evolution are
due to the thermal occupation numbers. that are static quantities.
Let us suppose that the system evolve with our effective Hamiltonian as found
for the thermodynamic behaviour. The Weyl-representation
\cite{CGTVV95} gives us an unified scheme for describing the
dynamical variables. The coupling constant $g$ rules the quantum
deviations from the harmonic behaviour and for vanishing $g$,
Eq.(\ref{morexpsimpl}) can be assumed to maintain its validity at
different real times, provided also that $\alpha$ is small enough;
the result is exact when $g\to 0$. For finite values of $g$, the
validity of this scheme involves also the amplitudes of the
Gaussian fluctuations ruled by the parameter $\alpha$. Therefore,
there is the same behaviour found for approaching static
correlators with the effective Hamiltonian \cite{CGTVV95}.

At this level, the averages in time of the quantities
$\langle\langle {\hat A}\rangle\rangle$ and $\langle\langle {\hat
B}(t)\rangle\rangle$, evolving with the effective Hamiltonian $\cal
H_{{}_{\rm eff}}(\xi)$ \cite{CaoVall}, can provide an approximation
for the time-dependent Mori product $({\hat A}|{\hat B}(t))$.

This procedure yields a good approximation for times up to the
order of $\hbar\beta$, for which the use of the effective potential
makes sense in the calculation of the static quantities at lowest
order, reproducing for instance a correct second moment for the
displacement-displacement dynamic correlator with a well-behaved
classical long time decay.


\begin{thebibliography}{10}

\bibitem{GTall}
R. Giachetti and V. Tognetti, Phys. Rev. Lett. {\bf 55},  912
(1985); Phys. Rev. B {\bf 33},  7647  (1986). R.~P. Feynman and H.
Kleinert, Phys. Rev. A {\bf 34},  5080 (1986).

\bibitem{CTVVmagall}
A. Cuccoli, V. Tognetti, P. Verrucchi, and R. Vaia, Phys. Rev. B {\bf 46},
  11601  (1992); Phys. Rev. Lett. {\bf 77},
  3439  (1996). E.~R. Cowley and G.~K. Horton,  in {\em Dynamic Properties of Solids,
  Vol. 7},
  edited by G.~K. Horton and A.~A. Maradudin (North-Holland, Amsterdam, 1995).

\bibitem{CGTVV95}
A. Cuccoli, R. Giachetti, V. Tognetti, R. Vaia and P. Verrucchi,
J. Phys. Cond. Matt. {\bf 7},  7891  (1995).

\bibitem{Moriboth}
H. Mori, Progr. Theor. Phys. {\bf 33},  423  (1965);
Prog. Theor. Phys. {\bf 34},  399  (1965).

\bibitem{CMGTVall}
A. Cuccoli, A. A. Maradudin, A. R. Mc Gurn, V. Tognetti and R. Vaia,
Phys. Rev. B {\bf 46},  8839  (1992); Phys. Rev. B {\bf 48},  7015  (1993).

\bibitem{LoveseyM72}
S.~W. Lovesey and R.~A. Meserve, J. Phys. C {\bf 6},  79  (1972);
M.H. Lee,J. Hong and J. Florencio, Phys. Scr. {\bf T 19}, 498,
1987.

\bibitem{CGTV98}
A. Cuccoli, R. Giachetti, V. Tognetti, R. Vaia, J.Phys. {\bf A31},
L419, 1998.


\bibitem{CaoVall}
J. Cao and G.~A. Voth, J. Chem. Phys. {\bf 99},  10070  (1993);
{\bf 100},  5093 and 5106, (1994).



\end{thebibliography}
\end{document}